\documentclass{emulateapj}

\slugcomment{Received 2008 November 27; accepted 2009 June 22; published 2009}

\shorttitle{Radio Variability in Seyferts}
\shortauthors{Mundell et al.}

\usepackage[]{natbib}
\begin{document}

\title{Radio Variability in Seyfert Nuclei}

\author{C.G. Mundell$^1$, P. Ferruit$^2$, N. Nagar$^3$ and A.S. Wilson$^4$}
\affil{$^1$~Astrophysics Research Institute, Liverpool John Moores
University, Twelve Quays House, Egerton Wharf, Birkenhead, CH41 1LD, U.K: cgm@astro.livjm.ac.uk}
\affil{$^2$~Observatoire de Lyon, 9 Avenue Charles Andr\'e,
Saint-Genis Laval Cedex, F69561, France}
\affil{$^3$~Astronomy Group, Departamento de F\'isica, Universidad de Concepci\'on, Casilla 160-C, Concepci\'on, Chile}
\affil{$^4$~University of Maryland, College Park, MD20742, U.S.A.}

\begin{abstract}
\noindent Comparison of 8.4-GHz radio images of a sample of eleven, early-type
 Seyfert galaxies with previous observations reveals possible variation in the nuclear radio flux density in five of them over a seven year period.  Four Seyferts (NGC~2110, NGC~3081,
 MCG~--6-30-15 and NGC~5273) show a decline in their 8.4-GHz nuclear
 flux density between 1992 and 1999, while one (NGC~4117) shows an
 increase; the flux densities of the remaining six Seyferts
 (Mrk~607, NGC~1386, Mrk~620, NGC~3516, NGC~4968 and NGC~7465) have
 remained constant over this period. New images of MCG~--5-23-16
 are also presented. We find no correlation between radio
 variability and nuclear radio luminosity or Seyfert nuclear type,
 although the sample is small and dominated by type 2
 Seyferts. Instead, a possible correlation between the presence of
 nuclear radio variability and the absence of hundred parsec-scale
 radio emission is seen, with four out of five marginally-resolved or
 unresolved nuclei showing a change in nuclear flux density, while
 five out of six extended sources show no nuclear variability despite
 having unresolved nuclear sources. NGC~2110 is the only source in our
 sample with significant extended radio structure and {\em strong} nuclear
 variability ($\sim$38\% decline in nuclear flux density over seven
 years). The observed nuclear flux variability indicates significant changes
are likely to have occurred in the structure of the nucleus on scales
smaller than the VLA beam size (i.e. within the central 15 pc), between
the two epochs, possibly due to the appearance and fading of new
components or shocks in the jet, consistent with previous detection of
sub-parsec scale nuclear structure in this Seyfert. Our results suggest that all
 Seyferts may exhibit variation in their nuclear radio flux density at
 8.4 GHz, but that variability is more easily recognised in compact
 sources in which emission from the variable nucleus is not diluted by
 unresolved, constant flux density radio-jet emission within the
 central $\lesssim$~50~pc.  If flares in  radio light curves
 correspond to ejection of new relativistic components or emergence of
 shocks in the underlying flow,  improved monitoring
 and high resolution imaging using VLBI techniques are required to confirm that radio
 jets are intrinsically non-relativistic during quiescence but that
 Seyferts, as black-hole driven AGN, have the capacity to accelerate
 relativistic jets during radio flares.  Finally, we conclude that our
results taken together with the
 increased detection rate of flat spectrum radio nuclei in Seyferts
 imaged at VLBI resolutions and the detection of variable water
 megamaser emission support the paradigm of intermittent periods of
 quiescence and nuclear outburst across the Seyfert population.
\end{abstract}

\section{INTRODUCTION}

Variability of nuclear flux density at all wavelengths across the
electromagnetic spectrum has long been recognized as a defining
characteristic of active galactic nuclei (AGN) \citep[e.g.][]{sh63}.
The short variability timescales measured at X-ray, optical and UV
wavelengths, ranging from days to years depending on observing
wavelength and intrinsic AGN luminosity, support the standard AGN
paradigm of nuclear emission originating from compact regions powered
by accretion of matter by a central supermassive black hole, rather
than standard stellar processes \citep[e.g.][]{um97,sa99,gl03}.

Variability at radio wavelengths is most marked in powerful radio-loud
AGN.  The largest and most rapid variations observed in nuclear
non-thermal continuum emission is seen in BL Lacs and core-dominated
quasars \citep[e.g.][and references therein]{li01}. These variations
are explained primarily by Doppler boosting of the nuclear emission by
highly-relativistic jets viewed at small angles to the line of sight
and are consistent with radio-loud Unification Schemes
\citep[e.g.][]{ob82}. Flares in the radio light curves might
correspond to ejection of new relativistic components or emergence of
shocks in the underlying flow \citep[e.g.][]{bh00,va02}. 

\setcounter{table}{0}
\begin{table*}[t]
\scriptsize
\caption{VLA observing parameters}
\label{tab:obsparams}
\begin{tabular}{lcccccrcc} \\
\hline\hline
Galaxy                         &	\hspace*{15mm}&	Phase&\hspace*{15mm}&Field  Center &\hspace*{12mm}&	Cycle time&\hspace*{10mm} & Time on Source \\
 	                           &	&Calibrator		&\hspace*{15mm}& RA ,	Dec (B1950)&\hspace*{12mm}&      (minutes)&\hspace*{10mm}& (minutes)\\
\hline
Mrk 607                      & &	0336$-$019		&&	03 22 17.77,	$-$03 13 04.8			&&	14.4~$+$~1.4			&&	140 \\
NGC 1386                   &&	0332$-$403		&&	03 34 51.80,	$-$36 09 47.1			&&	10.5~$+$~1.5			&&	160 \\
NGC 2110                   &  &	0539$-$057		&&	05 49 46.38,	$-$07 28 02.0			&& 	 9.5~$+$~1.5 			&&	130 \\
Mrk 620                      & &	0646$+$600		&&	06 45 37.63,	$+$60 54 12.8		&&	14.4~$+$~1.4			&&	140 \\
MCG$-$5$-$23$-$16& &	0919$-$260		&&	09 45 28.32,	$-$30 42 59.0			&&	14.6~$+$~1.6			&&	140 \\
NGC 3081                   & &	1032$-$199		&&	09 57 09.94,	$-$22 35 10.6			&&	14.5~$+$~1.5			&&	140 \\
NGC 3516                   & &	1044$+$719		&&	11 03 22.81,	$+$72 50 20.5		&&	10.4~$+$~1.4			&&	140 \\
NGC 4117                   & &	1144$+$402		&&	12 05 14.13,	$+$43 24 16.1		&&	14.5~$+$~1.5			&&	140 \\
NGC 4968                   & &	1256$-$220		&&	13 04 23.92,	$-$23 24 35.1			&&	14.5~$+$~1.5			&&	140 \\
MCG$-$6$-$30$-$15& &	1313$-$333		&&	13 33 01.84,	$-$34 02 25.7			& &	 9.4~$+$~1.4			&&	132 \\
NGC 5273                   & &	1328$+$307		&&	13 39 55.15,	$+$35 54 21.7		&&	14.5~$+$~1.5			&&	140 \\
NGC 7465                   & &	2251$+$158		&&	22 59 32.03,	$+$15 41 44.4		&&	14.5~$+$~1.5			&&	140 \\
\hline\\
\end{tabular}\\
\end{table*}

Seyfert nuclei, classified as radio quiet though not radio silent,
have been known for some time to show variability in their continuum
and line emission at X-ray, optical and UV wavelengths
\citep[e.g.][]{tu99,wa99,na00,pt00,sh01,mh05,wh08}, providing
constraints on the variation of nuclear absorbing column, photon
reprocessing, structure and dynamics of the broad line region and
ultimately the mass of the central object.  However, little is known
about radio variability in Seyfert nuclei as only a small number of
`interesting' Seyferts have been systematically monitored at radio
frequencies \citep[e.g.][]{nb83,wr00}, with some
studies motivated by serendipitous discovery of radio flares
\citep{fa98,fa00}.

Here we present the results of a two-epoch study of 8.4-GHz nuclear
radio emission in a small, but complete sample of 11 nearby
optically-selected Seyferts in which the nuclear radio emission is
imaged at angular resolutions of $\sim$0\farcs2 ($<$50 pc); the
original goal of the study was to obtain high quality radio images to
investigate the impact of Seyfert radio jets on the interstellar
medium by comparing extended radio structures with extended ionised
gas distributions and excitations in the narrow line region as
inferred from images with the {\em Hubble Space Telescope} ({\em HST}) by Ferruit,
Wilson \& Mulchaey (2000). We provide such a comparison here,
including new radio images of MCG~--5-23-16. However, by more reliably separating nuclear
and extended radio components and comparing with previous nuclear flux
density measurements we serendipitously discovered a
fraction of Seyfert nuclei with variable nuclear radio flux densities. It is
primarily this result that we present here.

VLA and MERLIN observations and data reduction are described in
Section 2 with the results of the VLA radio imaging, individual source
properties and comparison with HST emission-line maps presented in
Section 3. In Section 4, we discuss the nuclear radio variability and
its possible relationship to radio jet properties and the small number
of previously-studied  radio flares in Seyfert nuclei. Conclusions are
presented in Section 5. A Hubble constant of  H$_{\rm
0}$~=~75~km~s$^{-1}$~Mpc$^{-1}$ is used throughout.
  
\section{OBSERVATIONS AND DATA REDUCTION}
\subsection{Sample Selection}

We have selected a set of 12 Seyfert galaxies, which forms a
sub-sample of Mulchaey's sample of early-type Seyfert galaxies
\citep{mu96} and comprises all 12 Seyfert galaxies in early type
hosts with m$_{V}$~$<$~14.5, cz~$<$~3000 km s$^{-1}$ and declination
suitable for VLA observations. The optical continuum and ionized gas
properties have been studied with the {\em HST} \citep{fe00}; our sample is
the same as that of Ferruit et al. (2000), except that
two galaxies too far south to be reached by the VLA are omitted. All of our
objects, except MCG~--5-23-16, were observed with the VLA in 1993, at 1.4
GHz and 8.4 GHz, as part of the larger VLA snapshot study of Nagar et
al. (1999 - hereafter N99).

\subsection{VLA Observations}

The observations were conducted using the VLA in A configuration at
8.4 GHz on 1999 August 20 and 21 under project code AF360 (P.I.:
Ferruit). Each Seyfert was observed twice, in two $\sim$1-hr
segments separated in time to provide improved hour angle and $(u,v)$
coverage. Within each segment, the standard phase referencing
technique of cycling between the short exposures on the target and a
nearby, bright compact radio source (phase calibrator) was used in
order to correct for phase and gain variations due to instumental and
ionospheric variations. Observing parameters are given in Table
\ref{tab:obsparams} including the calibrator/target cycle time.  All
Seyferts, except NGC~3516, were observed in standard continuum mode
using two intermediate-frequency channels (IFs) resulting in a total
bandwidth of 100 MHz; dual circular polarizations (right and left)
were recorded for all sources, and only the parallel hands (i.e., RR
and LL) were correlated. In order to allow for better removal of
contaminating emission from a strong confusing source 4\farcm3 away
\citep[see][]{my92}, NGC~3516 was observed in spectral line mode,
using 7 channels and 2 IFs resulting in a total bandwidth of 37.5 MHz
after removal of end channels. Although NGC~3516 was observed for 140
minutes, data problems rendered 56 minutes of data unusable leaving 84
minutes of data available for imaging.  Observations of 0134+329
(3C48) and 1328+307 (3C286) were used to determine the flux density
scale using the VLA values as determined in 1995.2.

Standard data editing, calibration and analysis were performed
\citep{gm98}{\footnote{http://www.cv.nrao.edu/aips/cook.html} using
the NRAO\footnote{The National Radio Astronomy Observatory is a
facility of the National Science Foundation operated under cooperative
agreement by Associated Universities, Inc.}  Astronomical Image
Processing System \citep[AIPS;][]{vm96}. Data for each IF and
polarization were edited and calibrated separately before being
combined and imaged. Clean components from the resultant images were
then used to perform standard iterative self-calibration on the
stronger sources (S$_{8.4~GHz}$~$\gtrsim$~2~mJy) -- NGC~1386,
NGC~2110, Mrk~620, MCG~--5--23--16, and NGC~4968.

A similar but more extensive method was used for the weaker sources --
Mrk~607, NGC~3081, NGC~4117, MCG~--6-30-15 and NGC~5273.  Although it
was possible to image the weaker sources by applying directly the
interpolated phase and gain corrections derived from the phase
calibrator in the standard way, it was important to examine the data
for possible atmospheric decorrelation effects that could result in an
apparent reduction in flux density. A number of these weaker sources
were observed during the daytime in Summer when rapid phase changes at 8.4~GHz
due to variations in the troposphere might occur.  We therefore
inspected the raw phases observed for each phase calibrator and
calculated the rate of change of phase across each 1-minute scan and
within a 10-s integration time. Most scans showed small phase changes
$\lesssim$10$-$20$^{\circ}$ per 10~s integration, resulting in
amplitude reduction due to decorrelation of less than 6\%. A maximum
phase rate of $\sim$4$^{\circ}$~s$^{-1}$ was observed on some of the
longest baselines for $<$30\% of the time; the reduction in amplitude
from decorrelation at this phase rate would be $\sim$22\% so data
corresponding to these timeranges were excised. Assuming phase changes
monotomically across a 1-minute scan (i.e. that extreme 2$\pi$ phase
wraps do not occur between integrations), the resultant maximum phase
rates per 10-s integration for the remaining data are
negligible. Atmospheric decorrelation within an integration time is
therefore ruled out.

As each Seyfert was observed in two 1-hr segments, we used the
observed phase rates for the associated phase calibrator to quantify
the data quality for each 1-hr data segment separately. In general,
each phase calibrator has very good phase stability for at least 1
hr where `good' is defined as negligible phase rates within each
1-minute calibrator scan and only small, smoothly varying, phase rates
between scans on most baselines.  With this quality definition in
mind, each hour segment was imaged separately and the final flux
densities measured. As expected, the images resulting from the time
range with poorer phase stability show a reduced central flux density
of up to $\sim$30\% compared with the `good' timerange, with the
remaining flux density spread throughout the map. Therefore, only
clean components from the `good' image were used to perform phase-only
self-calibration on the whole 2-hr dataset. Phases were re-examined
and each 1-hr segment re-imaged separately. In all cases the
self-calibration worked well for the `good' hour and in the case of
MCG~--6-30-15 it also worked well for the `poor' hour, as confirmed by
the flux density in each resultant image being the same. A final image
combining the full two hours was then made. In the case of NGC~3081,
Mrk~607, NGC~4117 and NGC5273, the self-calibration of the `poor' hour
was insufficiently good to warrant inclusion of these data in the
final image and only the `good' hour was used. This resulted in a
higher rms noise in the final images due to the reduced exposure
time, but gives a more reliable total flux density estimate, free from
significant phase decorrelation effects.

As described above, NGC~3516 was observed in spectral line mode;
therefore, in addition to editing and calibrating each IF and
polarization separately, each narrow spectral line channel was also
calibrated, editted, and imaged separately before being averaged
together to form the final image ready for deconvolution. This method
minimizes chromatic aberration, or bandwidth smearing, of the
confusing source and ensures correct removal of its sidelobe
contribution at the phase center of the image close to NGC~3516. A
Gaussian fit to the confusing source confirmed that it was unresolved
and not radially smeared, therefore free from chromatic aberration.
Iterative self-calibration, using the same imaging and calibration
method preserving the narrow spectral channels until the last stage of
imaging, was then performed on the spectral line dataset,
using clean components from the final combined image.

The final calibrated $(u,v)$ data for all sources were Fourier
transformed with both natural and uniform weighting and deconvolved;
the highest angular resolution images were made using uniform
weighting with Briggs' robustness parameter set to $-$4 (equivalent to
uniform weighting) while lower resolution images, more sensitive to
extended emission, were made with both natural weighting and uniform
weighting with robust paramenter 0.  Final images (Figures
\ref{fig:Seyferts1}-\ref{fig:Seyferts4}) were restored with circular
beams, with sizes corresponding to the mean value of the full width at
half maximum of minor and major axes of the fitted beam. The data were
also tapered (reducing the contribution from data points corresponding
to longer baselines) to search for more extended emission; a tapered
image of NGC~3516 is shown in Figure \ref{fig:tap3516}.
Comparison of parameters derived from Gaussian fits to images restored
with a fitted elliptical beam and the mean circular beam showed no
difference, indicating that for most sources a circular beam is a
reasonable choice. The source properties for NGC~3516 (Table
\ref{tab:components}) were derived from an image (not shown) restored
with the fitted elliptical beam which has its major axis in
P.A.~90$^{\circ}$ and thus maximum resolution in the north-south
direction, allowing the northern component to be fully separated from
the nuclear component. The integrated flux densities (Table
\ref{tab:allgalprops}) were measured from the naturally weighted
images, which are most sensitive to extended structure, and individual component properties (Table
\ref{tab:components}) from uniformly weighted images, which provide maximum angular resoluion. The uncertainty
in the flux scale is taken to be 5\% and is included in the total
uncertainties in flux densities quoted in Tables \ref{tab:allgalprops}
and \ref{tab:components}. These errors were derived by adding, in
quadrature, the conservative 5\% flux scale error, the rms noise in
the final image and the error in the Gaussian fitting. Contour levels,
as multiples of 3-$\sigma$ rms noise, and restoring beam sizes used
in Figures \ref{fig:Seyferts1}~-~\ref{fig:Seyferts4} are listed in
Table \ref{tab:contlevs}. 

The original data presented in N99 were reprocessed using the methods described above to provide error estimates, not originally quoted in N99, test for any decorrelations and to enable consistent flux comparisons across the two epochs. The N99 data were obtained during winter months when decorrelation is less severe, consistent with our tests; we verified this in particular for NGC~4117, thereby ruling out decorrelation as an explanation for its non-detection in 1992. All sources were observed in A configuration,  except NGC~1386 and MCG~--6-30-15 for which lower angular resolution data from BnA configuration were obtained. Source values for both epochs are given in Tables \ref{tab:allgalprops} and \ref{tab:components}.

 \begin{figure*}
   \centering
   \includegraphics[width=12.5cm]{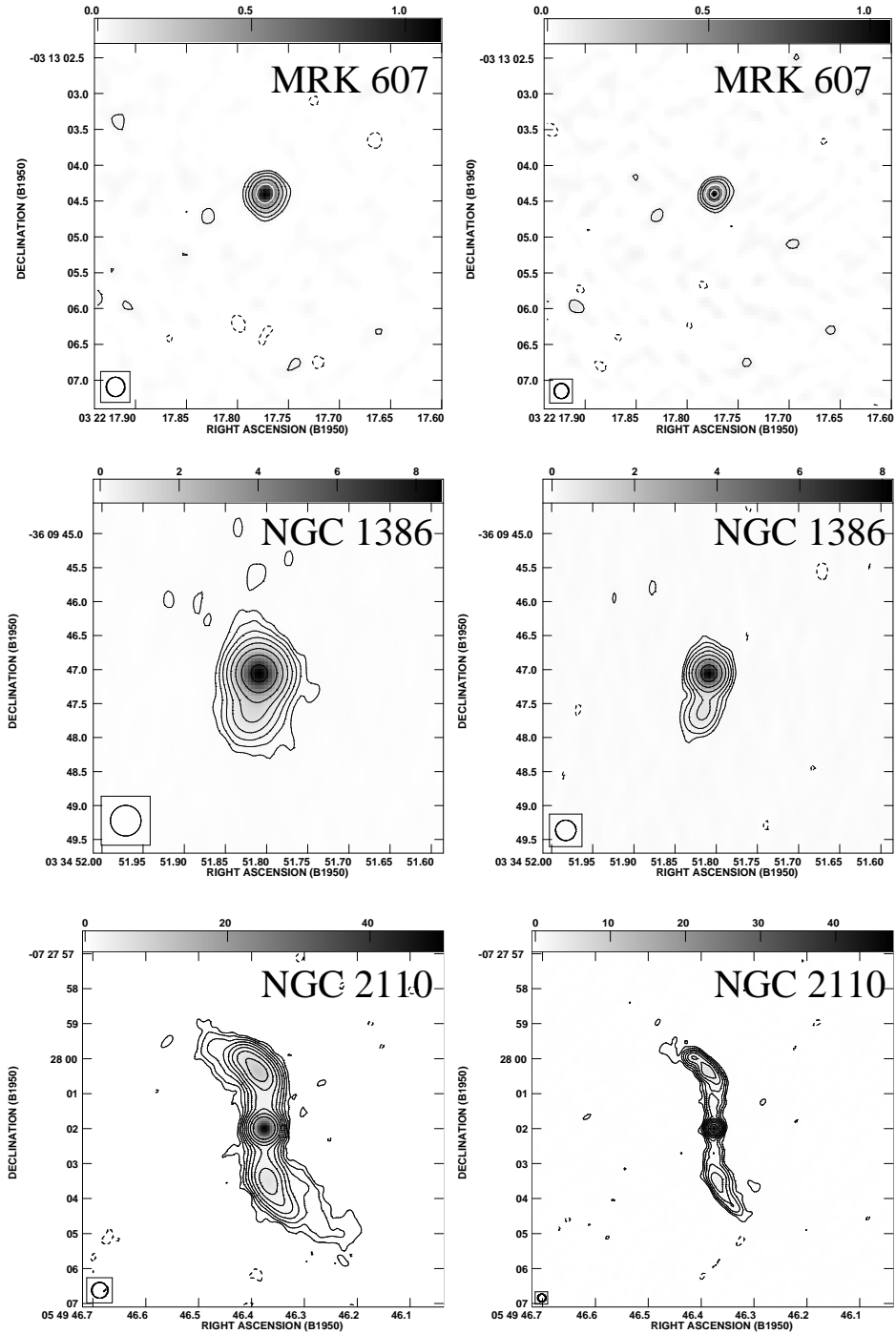}
   \caption{8.4-GHz radio images of Seyferts - lower/higher
resolution images are shown in the left/right column respectively. The restoring beam size is plotted in the lower
left corner of each image. Table \ref{tab:contlevs} gives details of restoring beam sizes and
contour levels used.}
              \label{fig:Seyferts1}
    \end{figure*}

\subsection{MERLIN Observations}

We obtained and analyzed two hours of MERLIN $\lambda$21-cm (1.407
GHz) archival data for Mrk~620, which was observed on 1998  May 9 with
seven antennas including the Lovell telescope. The data were processed
using standard methods \citep[e.g.][]{mn01},
including self-calibration and re-weighting of the antennas according
to their gains.
The presence of weak extended emission on the shorter spacings and the
small amount of data (restricted $(u,v)$ coverage) made
self-calibration and reliable imaging of weaker features difficult;
nevertheless flux densities of the brighter eastern and western
components, measured from a 0\farcs2$-$resolution image, were used to
calculate the spectral index between 1.4 GHz and 8.4 GHz at this
resolution.

\section{RESULTS}

\begin{figure*}
\centering
   \includegraphics[width=12.5cm]{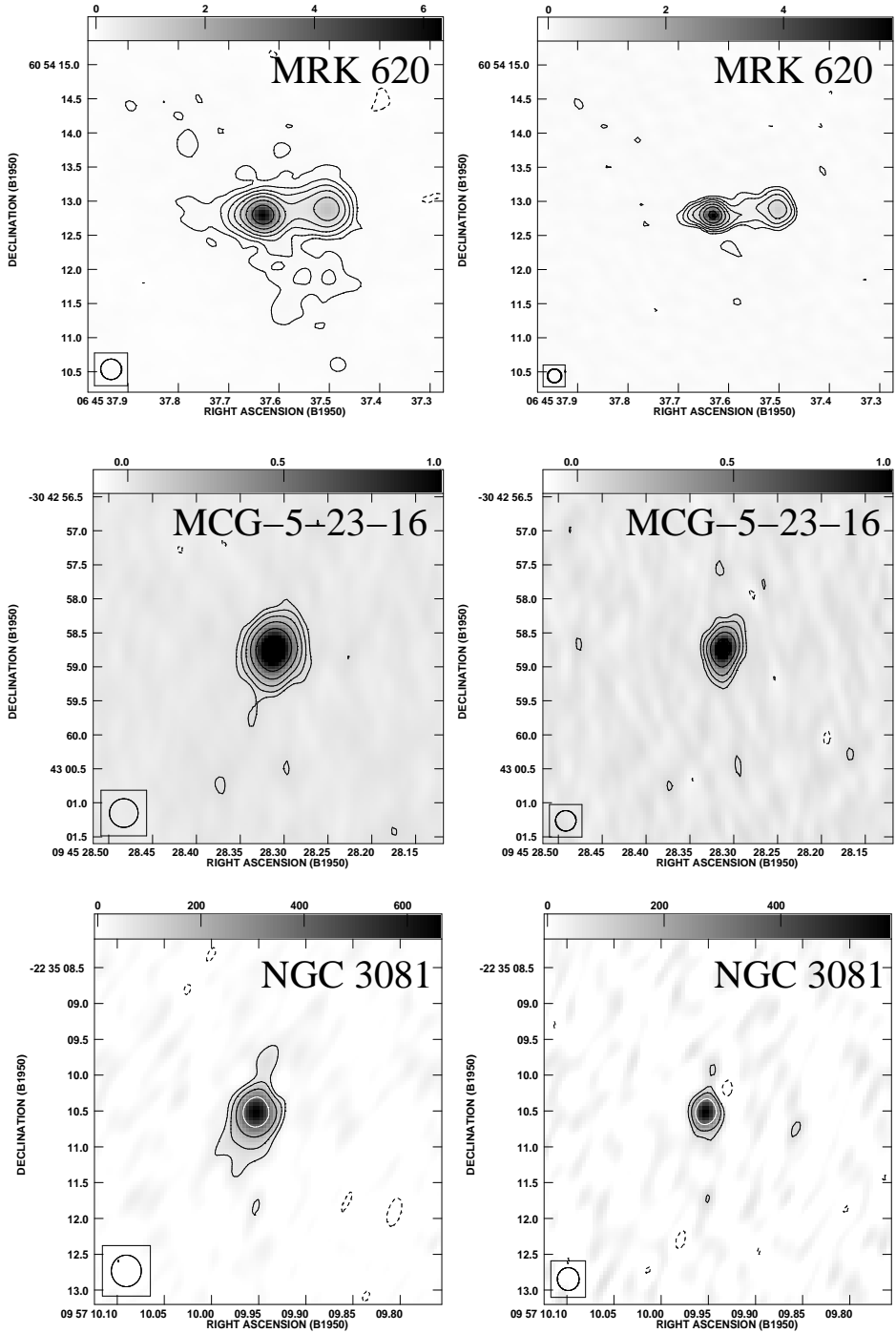}
\caption{8.4-GHz radio images of Seyferts; see Table \ref{tab:contlevs} for
contour levels.}
              \label{fig:Seyferts2}
    \end{figure*}

All 12 Seyfert nuclei in our sample are detected at 8.4 GHz with the VLA. For
the eleven Seyferts with two epochs of observations, four nuclei,
NGC~2110, NGC~3081, MCG~--6-30-15 and NGC~5273, show a decline in their
nuclear 8.4 GHz flux density between 1992 and 1999, while NGC~4117
shows an increase; the flux densities of the remaining six Seyferts,
Mrk~607, NGC~1386, Mrk~620, NGC~3516, NGC~4968 and NGC~7465, have
remained constant over this period. Here we describe the radio
properties of each Seyfert galaxy, comparing the present data with
earlier 8.4-GHz measurements by N99; sources with spatially extended
radio emission are compared with their optical line-emission images.

\subsection{Notes on Individual Sources}

\begin{figure*}
   \centering
   \includegraphics[width=12.5cm]{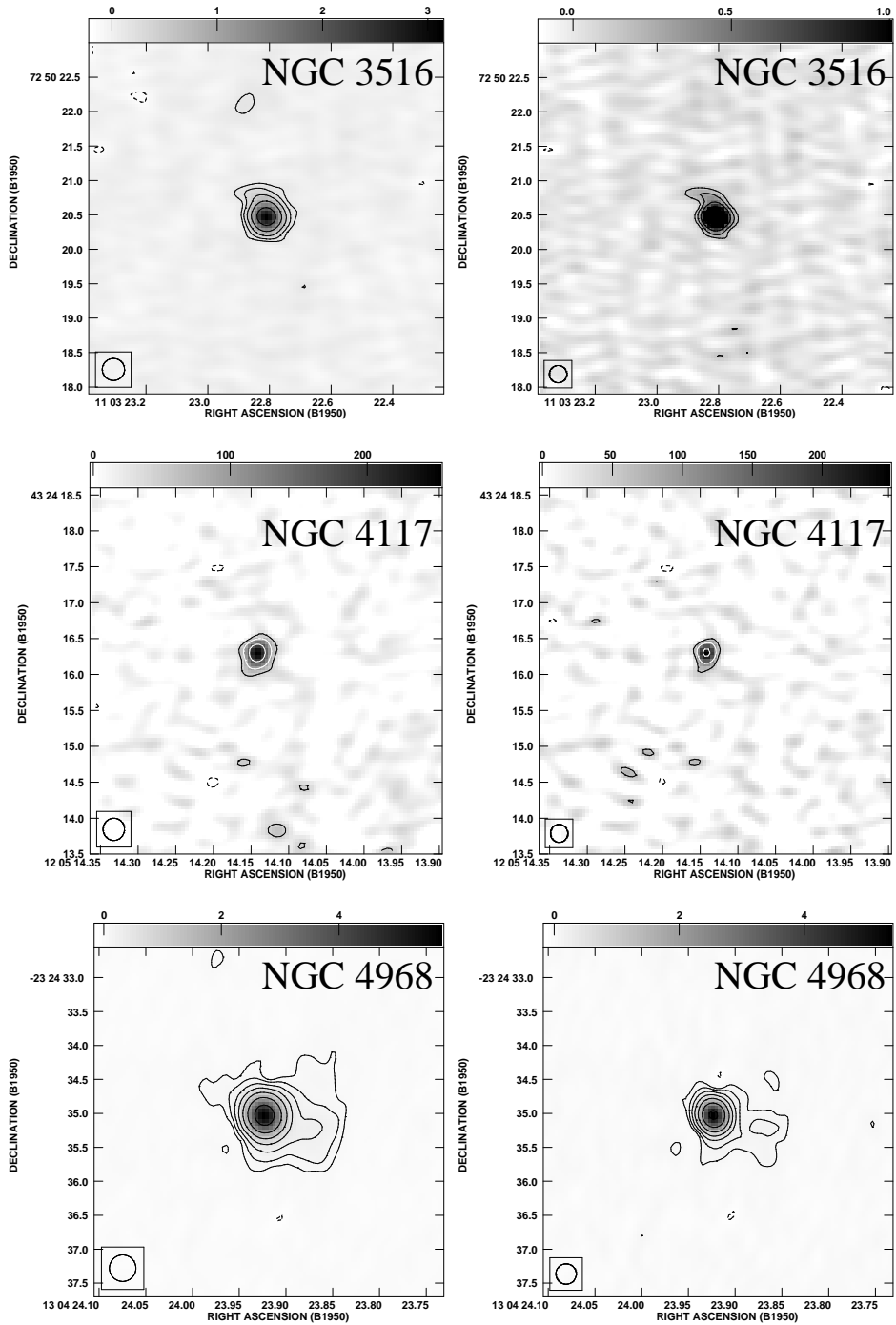}
  \caption{8.4-GHz radio images of Seyferts; see Table \ref{tab:contlevs} for
contour levels.}
              \label{fig:Seyferts3}
    \end{figure*}

{\em {\object Mrk~607} (NGC1320) - Seyfert type 2} - We confirm the presence of
a small, weak extension to the south of the nucleus, visible in the
naturally weighted image (Figure \ref{fig:Seyferts1}), which was previously
noted by N99 but presumed to be an
artefact; Gaussian fitting to the image suggests the source is
marginally resolved in PA 158$^{\circ}$ with a deconvolved size of
140~mas, although this value is uncertain due to the weakness of the
extended feature. The slightly higher flux density of 1.3 mJy measured
in the natural image compared with 1.2 mJy measured in the uniform
image (Figure \ref{fig:Seyferts1}) also suggests the possible presence of
weak extended emission. The total flux density of
Mrk~607 measured in 1999 remains the same as that measured in 1992 at
1.3 mJy.

Emission-line images \citep{fe00} show a ridge
of ionised gas extending up to 2$''$ northwest of the nucleus, along
the galaxy major axis, with the [O {\sc iii}]/[N {\sc ii}]+H$\alpha$
ratio peaking at the nucleus. The anti-correlation between the
extended radio emission and ionised gas, the high excitation level of
the nuclear optical line emission and the small linewidths of low and
medium excitation lines \citep{do86} suggest that the line-emitting
gas is photo-ionised by the nucleus and not by the passage of the
radio jet.

\begin{figure*}
   \centering
   \includegraphics[width=12.5cm]{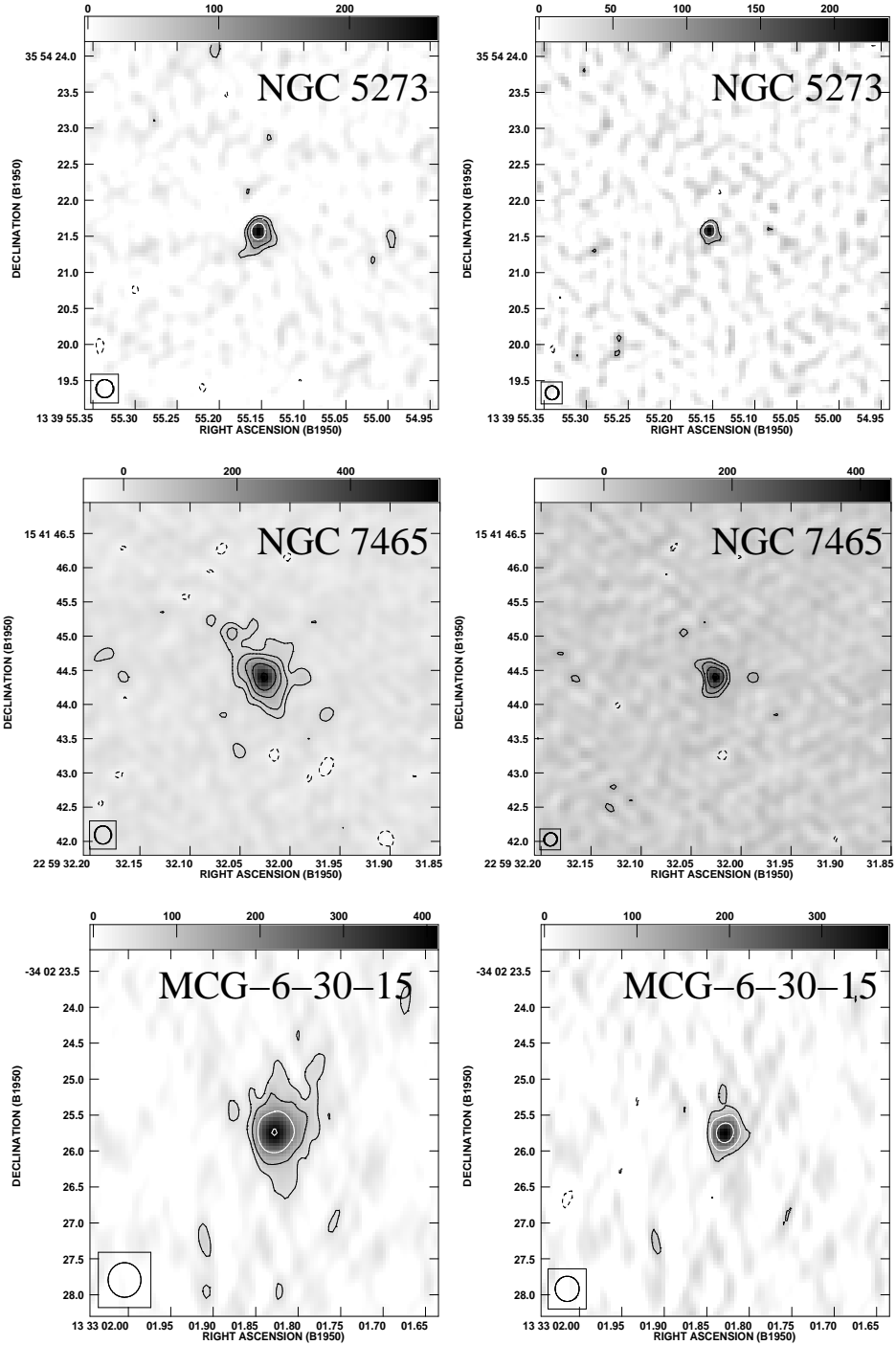}
  \caption{8.4-GHz radio images of Seyferts; see Table \ref{tab:contlevs} for
contour levels.}
              \label{fig:Seyferts4}
    \end{figure*}

\begin{table*}
\scriptsize
\caption{Results of 8.4 GHz observations and
comparison with previous observations}
\label{tab:allgalprops}
\begin{tabular}{lccccccrrccc}
\hline\hline
Galaxy     &\hspace{5mm} &Other&\hspace{3.5mm} &		R.A. &	Dec.&\hspace{3.5mm} &	S (mJy)& 	S (mJy)&Radio&Variable& Seyfert\\
		&	                        &         & &(B1950)	&(B1950)		&  &(1999)& 	(1992)&Structure&nucleus?&type\\
\hline
Mrk~607&&	NGC~1320&&	03 22 17.77&	$-$03 13 04.4&&	1.3$\pm$0.07&	1.3$\pm$0.2&	(S)&	No&2\\
NGC~1386&&	ESO 358$-$G35&&	03 34 51.81&	$-$36 09 47.1&&	10.3$\pm$0.5&	9.1$\pm$0.6&	L&	No&2\\
NGC~2110&&	MCG$-$1$-$15$-$4&&05 49 46.38&	$-$07 28 02.0&&	101.6 $\pm$5.1&	130.9$\pm$6.5&	L&	Decline&2\\
Mrk~620&&	NGC~2273&&	06 45 37.63&	$+$60 54 12.8&&	13.5$\pm$0.7&	11.4$\pm$0.6&	L$+$D&	No &2\\
MCG$-$5$-$23$-$16&&ESO 434$-$G40&&09 45 28.31&	$-$30 42 58.8&&	2.1 $\pm$0.1&	$-$&	(S)&	$-$&2 \\
NGC~3081&&	ESO 499$-$IG31&&	09 57 09.95&	$-$22 35 10.6&&	0.72$\pm$0.05&	1.35$\pm$0.1&	(S)&	Decline&2\\
NGC~3516&&	UGC~6153&&	11 03 22.81&	$+$72 50 20.5&&	3.9$\pm$0.21&	4.1$\pm$0.2&	L&	No&1.2\\
NGC~4117&&	UGC~7112&&	12 05 14.14&	$+$43 24 16.3&&	0.3$\pm$0.05&	$<$0.1&	U&	Increase&2 \\
NGC~4968&&	MCG$-$4$-$31$-$30&&13 04 23.92&	$-$23 24 35.0&&	7.8$\pm$0.7&	7.4$\pm$0.4&	S$+$D&	No&2\\
MCG$-$6$-$30$-$15&&ESO 383$-$G35&&13 33 01.83&	$-$34 02 25.7&&	0.64$\pm$0.05&	1.0$\pm$0.1&	(S)&	Decline&1.2\\
NGC~5273&&	UGC~8675&&	13 39 55.15&	 $+$35 54 21.6&&	0.38$\pm$0.06&	0.64$\pm$0.08&	(S)&	Decline&1.5\\
NGC~7465&&	UGC~12317&&	22 59 32.03&	 $+$15 41 44.4&&	1.2$\pm$0.1&	1.4$\pm$0.1&	S&	No&2\\
\hline\hline 
\end{tabular}\\\\
{\bf Notes.} Galaxy names, alternative names, positions and total integrated fluxes
measured in 1999 are listed (columns [1]$-$[5]). Listed for comparison
(column [6]) are the total integrated fluxes measured in 1992 by Nagar
et al. (1999) corrected by a factor of 1.006 to correct for the small
difference in absolute flux scale calibration adopted by N99 and the
present paper. The radio source structure (Column [7]) in the 8.4 GHz
images is listed, following the classification definitions of Ulvestad
\& Wilson (1984a) where L~=~linear source, S~=~slightly resolved,
D~=~diffuse and U~=~unresolved; sources which are less than one half
the beam size after deconvolution but show a possible extension in the
3.6 cm image are tentatively considered to be slightly resolved
(S). Any variability in flux density of the nuclear component between
the 1992 and 1999 observations is noted in column [8], but see text
for full discussion of source properties; see also Table
\ref{tab:components} for properties of multiple-component resolved
sources. MCG$-$5$-$23$-$16 was not observed by N99, so no flux 
comparison is made here, as indicated by ``$-$'' in columns (6) and (8).\\
\end{table*}

\begin{table*}
\scriptsize
\caption{Resolved sources - component properties.}
\label{tab:components}
\begin{tabular}{lcccccrccccc}
\hline\hline
Galaxy&\hspace{5mm} &R.A. (B1950)&\hspace{5mm} &	Dec. (B1950)&\hspace{5mm} &S (mJy)('99)&\hspace{3mm} &	S  (mJy)('92)&\hspace{5mm} &Component&	Variability?\\
\hline
NGC~1386&&	03 34 51.81&&$-$36 09 47.1&&	9.0$\pm$0.5&&	9.1$\pm$0.6 total&&	Core&		No \\
NGC~1386&&	03 34 51.82&&$-$36 09 47.6&&	1.3$\pm$0.1&&	$-$&&		South  &	$-$\\
NGC~2110&&	05 49 46.38&&$-$07 28 02.0&&	50.8 $\pm$2.5&&	81.7$\pm$4.1&&		Core&		Decline\\
NGC~2110&&	05 49 46.39&&$-$07 28 00.4&&	32.1$\pm$1.6&&	33.4$\pm$1.7&&		North Jet &	No\\
NGC~2110&&	05 49 46.37&&$-$07 28 03.5&&	16.7 $\pm$1.1&&	15.8$\pm$1.1&&		South Jet&	No\\
Mrk~620&&	06 45 37.63&&$+$60 54 12.8&&	7.1$\pm$0.4&&	6.9$\pm$0.4&&		Core (East)&	No \\
Mrk~620&&	06 45 37.50&&$+$60 54 12.9&&	2.5$\pm$0.2&&	2.9$\pm$0.4&&		West 	&	No\\
Mrk~620&&		$-$     &&    $-$              &&	$\sim$3.9    &&	       $-$&&		Ext. NE \& SW&	$-$\\
NGC~3516&&	11 03 22.81&&$+$72 50 20.5&&	2.5$\pm$0.17&&	4.1$\pm$0.2 total&&	Core	&	No\\
NGC~3516&&	11 03 22.84&&$+$72 50 20.8&&	1.4$\pm$0.23&&	$-$&&		North  &	$-$\\
NGC~4968&&	13 04 23.92&&$-$23 24 35.0&&	6.1$\pm$0.4&&	7.4$\pm$0.4 total&&		Core	&	No\\
NGC~4968&&	13 04 23.88&&$-$23 24 35.2&&	1.7$\pm$0.6&&	$-$&&		West 	&	$-$\\
NGC~7465&&	22 59 32.03&&$+$15 41 44.4&&	$\sim$0.8&&	1.4$\pm$0.1 total&&	Core	&	No\\
NGC~7465&&	    $-$     &&	  $-$       &&	$\sim$0.4&&	$-$&&		NW extension&	$-$\\
\hline 
\end{tabular}\\
{\bf Note.} Galaxy name (column [1]), radio component position (column [2]), radio
component flux density in 1992 and 1999 (columns [4],[5]), component
identification (column [6]), variability between 1992 and 1999 (column
[7]).
\end{table*}
\normalsize

{\em {\object NGC~1386} - Seyfert type 2} - A clear extension to the south of the
nucleus is seen in both natural and uniform radio images (Figure
\ref{fig:Seyferts1}), confirming the detection by N99. The low declination
of the source results in a beam elongated in the north-south
direction, i.e. approximately along the source extension
direction. Nevertheless, the source is sufficiently bright for an
average circular restoring beam to show the extension clearly. Fitted
parameters are consistent with those derived from a uniform image with
elliptical beam (not shown). The unresolved nuclear flux density is
9.0 mJy and the southern component has a flux density of 1.3 mJy with
its peak lying $\sim$0\farcs52 south of the nucleus in PA
$\sim$175$^{\circ}$. Extended emission is also marginally resolved to
the north of the nucleus as can be seen in Figure \ref{fig:1386sub},
which shows the naturally-weighted residual radio image of NGC~1386
after subtraction, in the $(u,v)$ plane, of a 7.3 mJy point source centred
at the position marked ``+''. It is unlikely that there has been any
variation in nuclear flux density between 1992 and 1999 (see Table
\ref{tab:components}) and the higher total flux density in 1999 is
probably a result of the more reliable imaging of the extended component in the
present data. As shown in Table \ref{tab:components} the nuclear flux
density is 9.0 mJy, with an additional 1.3 mJy in the southern
extended component. 

Ionised gas lies in a string of emission-line knots extending more
than 2$''$ (180 pc) north and south of the nucleus in a direction
similar to that of the extended radio emission. However, comparison of
the [O {\sc iii}] and radio images (Figure \ref{fig:1386sub}) shows no
direct association between radio and optical components. 

{\em {\object NGC~2110} (type 2)} - The previously known
\citep{uw83,uw84,N99} symmetrical curved jet-like features extending
north and south of the nucleus are clearly visible in Figure
\ref{fig:Seyferts1}. The fainter, inner 1$''$ (150 pc) of the radio jets
appear straight and well collimated before changing direction at the
brighter hot-spots $\sim$1\farcs5 from the nucleus. In addition, the
lower resolution, naturally weighted image reveals a weaker, more
diffuse component extending away from each jet end, most noticable in
the south-west, possibly indicating disruption and de-collimation of
the radio jets or an extended radio cocoon flowing along the galactic
density gradient.  We do not detect any significant extended nuclear
emission perpendicular to the radio jets and conclude that the small
extension to the east of the nucleus seen by N99 is probably an
imaging artefact due to the limited $(u,v)$ coverage of their snapshot
observations.

The flux density of nucleus has declined by 30.9 mJy or $\sim$38\% in
seven years, while the flux density of the radio jets, has, as
expected, remained constant over this period, increasing our
confidence in the variability measurement of the nucleus. NGC~2110 is
the only Seyfert in this sample with significant {\em extended} radio
emission and a variable nucleus suggesting that the value of
$\sim$38\% variability is a lower limit; detection of such a variation
in flux density, despite contamination by unresolved jet emission,
implies very strong radio flux variability in the inner $<$0\farcs1
($\sim$15 pc).

The relationship between the ionised gas and radio emission has been
investigated by Mulchaey et al. (1994), Ferruit et al. (1998), and
Ferruit et al. (2004).

 \begin{figure*}
   \centering
   \includegraphics[width=15.5cm]{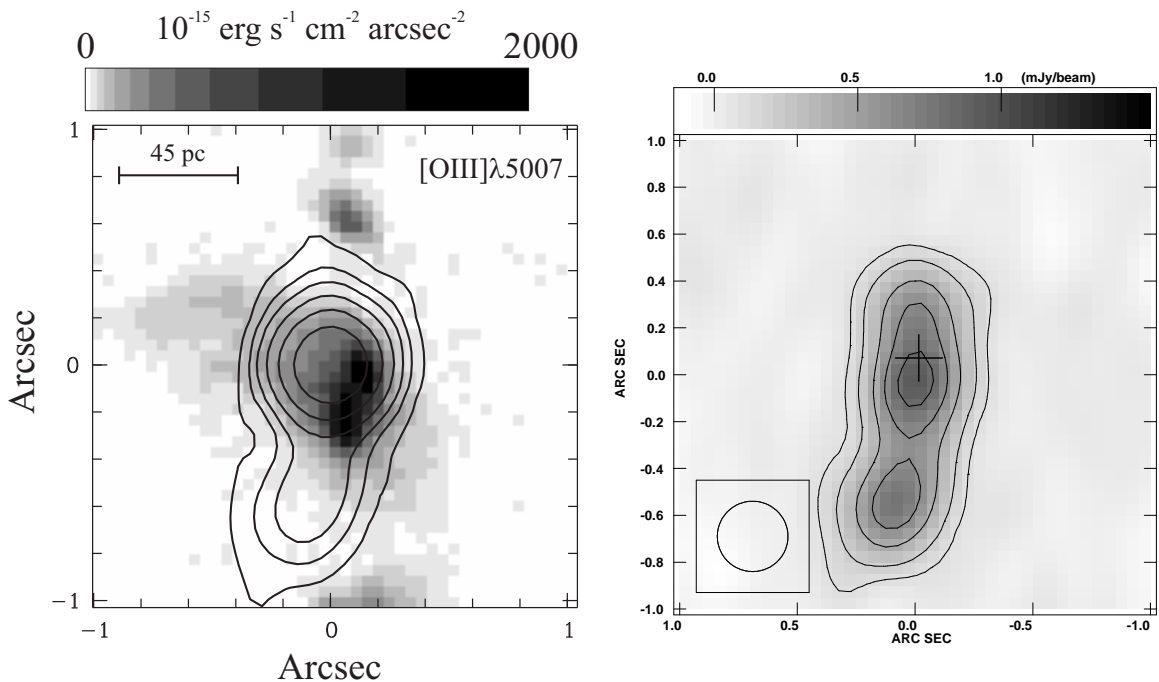}
\vspace*{-7mm}
\caption{Left: 8.4-GHz radio continuum image of  NGC~1386 (contours)
overlaid on the {\em HST} [O {\sc iii}]$\lambda$5007~\AA image from Ferruit
et al. (2000) (grey). The range in [O {\sc iii}] intensity, as indicated by the
grey-scale bar above the image, is 0$-$2$\times$10$^{-12}$
ergs~s$^{-1}$~ cm$^{-2}$~arcsec$^{-2}$. Contour levels for the 8.4-GHz
radio emission are (1, 3, 6, 12, 24, 48)*83~$\mu$Jy~beam$^{-1}$; Right:
8.4-GHz residual radio image of NGC~1386 after subtraction of a point
source of flux 7.3 mJy; the position of the subtracted
point source is marked ``+''. Contour levels are (-1, 1, 2, 4, 6,
8)*0.1~mJy~beam$^{-1}$}
\label{fig:1386sub}
\vspace{-9.9mm}
\end{figure*}

  \begin{figure*}
   \centering
   \includegraphics[width=15.6cm]{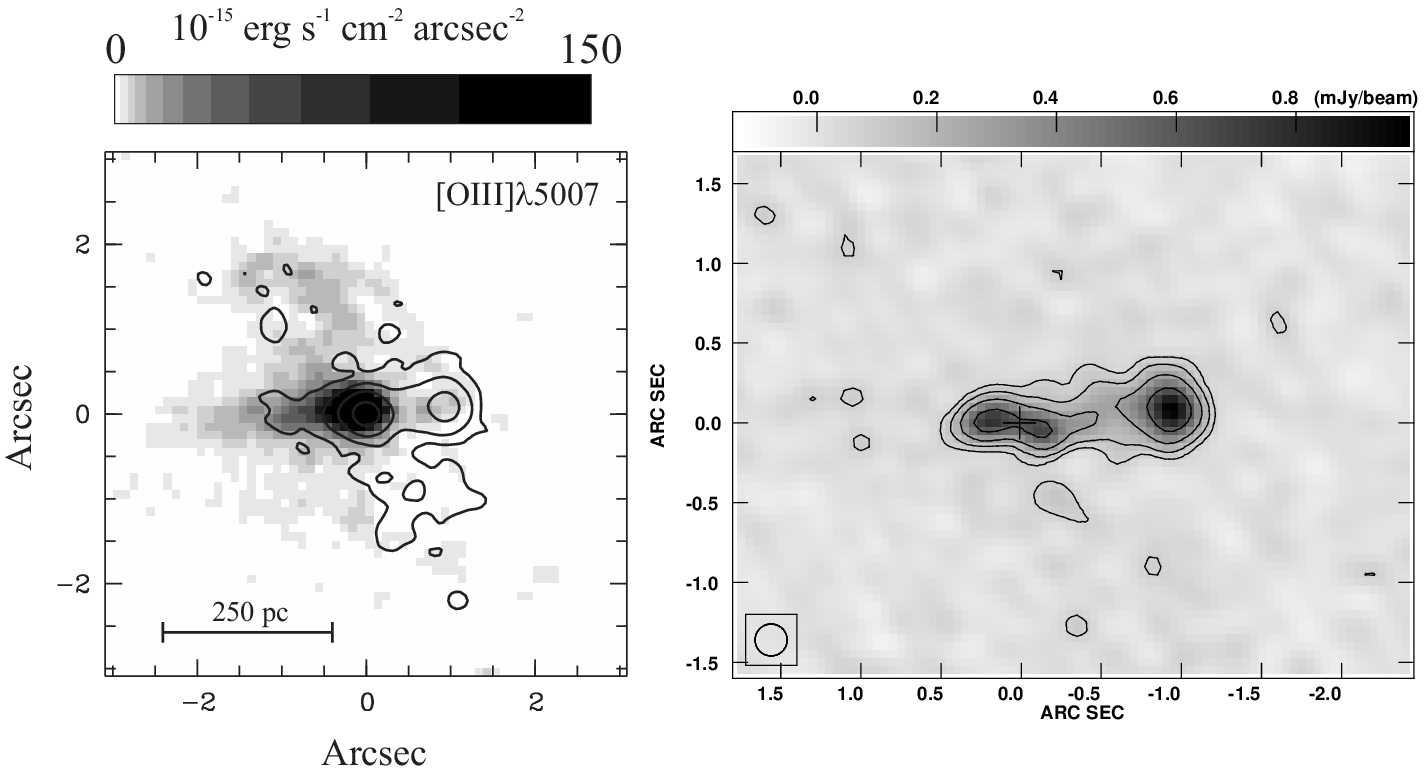}
\vspace*{-7mm}
\caption{Left: 8.4-GHz radio continuum image of Mrk~620 (contours) overlaid on the {\em HST}  [O {\sc iii}]$\lambda$5007~\AA\ image from Ferruit et al. (2000) (grey). The range in [O {\sc iii}] intensity, as indicated by the grey-scale bar above the image, is  0$-$1.5$\times$10$^{-13}$ ergs~s$^{-1}$~cm$^{-2}$~arcsec$^{-2}$. Contour levels for the 8.4-GHz radio emission are (1, 4, 16, 64)*55.2 $\mu$Jy~beam$^{-1}$; Right: 8.4-GHz residual radio image of Mrk 620, after subtraction of a point source of flux 7.1 mJy; the position of the subtracted point source is marked with a ``+''. Contour levels are (-1, 1, 2, 4, 8)*71.6~$\mu$Jy~beam$^{-1}$.}
\label{fig:m620sub}
\end{figure*}

\begin{table}
\scriptsize
\caption{Parameters for contour plots in Figures
\ref{fig:Seyferts1}~-~\ref{fig:Seyferts4}.}
\label{tab:contlevs}
\begin{tabular}{lccc} \\
\hline
Galaxy		&Beam Size &3~$\times$~rms &\hspace{-3mm}Contour levels \\
		&(arcsec) &($\mu$Jy beam$^{-1}$)&\hspace{-3mm}(in multiples of 3$\times$rms)\\
\hline
&&{\em Figure \ref{fig:Seyferts1}}&\\
Mrk 607		&0.27$\times$0.27& 39.0&\hspace{-3mm} (-1, 1, 2, 4, 8,16)\\
		&0.21$\times$0.21& 54.0	&\hspace{-3mm} (-1,1,2,4,8,16)\\
NGC 1386	&0.45$\times$0.45& 54.7	&\hspace{-3mm} (-1,1,2,4,8,16,32,64,128)\\
		&0.30$\times$0.30& 83.0	&\hspace{-3mm} (-1,1,2,4,8,16,32,64)\\
NGC 2110	&0.45$\times$0.45& 54.2	&\hspace{-3mm} (-1,1,2,4,8,16,32,64,128,256)\\
		&0.22$\times$0.22& 98.0	&\hspace{-3mm} (-1,1,2,4,8,16,32,64,128,256)\\
&&{\em Figure \ref{fig:Seyferts2}}&\\
Mrk 620		&0.30$\times$0.30& 55.2	&\hspace{-3mm} (-1,1,2,4,8,16,32,64)\\
		&0.20$\times$0.20& 71.4	&\hspace{-3mm} (-1,1,2,4,8,16,32,64)\\
MCG-5-23-16&0.42$\times$0.42& 37.1	&\hspace{-3mm} (-1,1,2,4,8,16,32)\\
		&0.30$\times$0.30& 60.0	&\hspace{-3mm} (-1,1,2,4,8,16)\\
NGC 3081	&0.44$\times$0.44& 50.0	&\hspace{-3mm} (-1,1,2,4,8)\\
		&0.32$\times$0.32& 75.0	&\hspace{-3mm} (-1,1,2,4)\\
&&{\em Figure \ref{fig:Seyferts3}}&\\
NGC 3516	&0.32$\times$0.32& 125	&\hspace{-3mm} (-1,1,2,4,8,16)\\
		&0.25$\times$0.25& 144	&\hspace{-3mm} (-1,1,2,4,8,16)\\
NGC 4117	&0.31$\times$0.31& 43.3	&\hspace{-3mm} (-1,1,2,4)\\
		&0.23$\times$0.25& 55.2	&\hspace{-3mm} (-1,1,2,4)\\
NGC 4968	&0.39$\times$0.39& 63.0	&\hspace{-3mm} (-1,1,2,4,8,16,32,64)\\
		&0.30$\times$0.30& 75.1	&\hspace{-3mm} (-1,1,2,4,8,16,32,64)\\
&&{\em Figure \ref{fig:Seyferts4}}&\\
NGC 5273	&0.25$\times$0.25& 42.1	&\hspace{-3mm} (-1,1,2,4)\\
		&0.19$\times$0.19& 66.0	&\hspace{-3mm} (-1,1,2)\\
NGC 7465	&0.26$\times$0.24& 31.8	&\hspace{-3mm} (-1,1,2,4,8,16)\\
		&0.19$\times$0.18& 48.2	&\hspace{-3mm} (-1,1,2,4,8)\\
MCG-6-30-15&0.48$\times$0.48& 50.0	&\hspace{-3mm} (-1,1,2,4,8)\\
		&0.35$\times$0.35& 63.0	&\hspace{-3mm} (-1,1,2,4)\\
\hline
\end{tabular}\\
\end{table}

   \begin{figure} 
\centering
   \includegraphics[width=8.5cm]{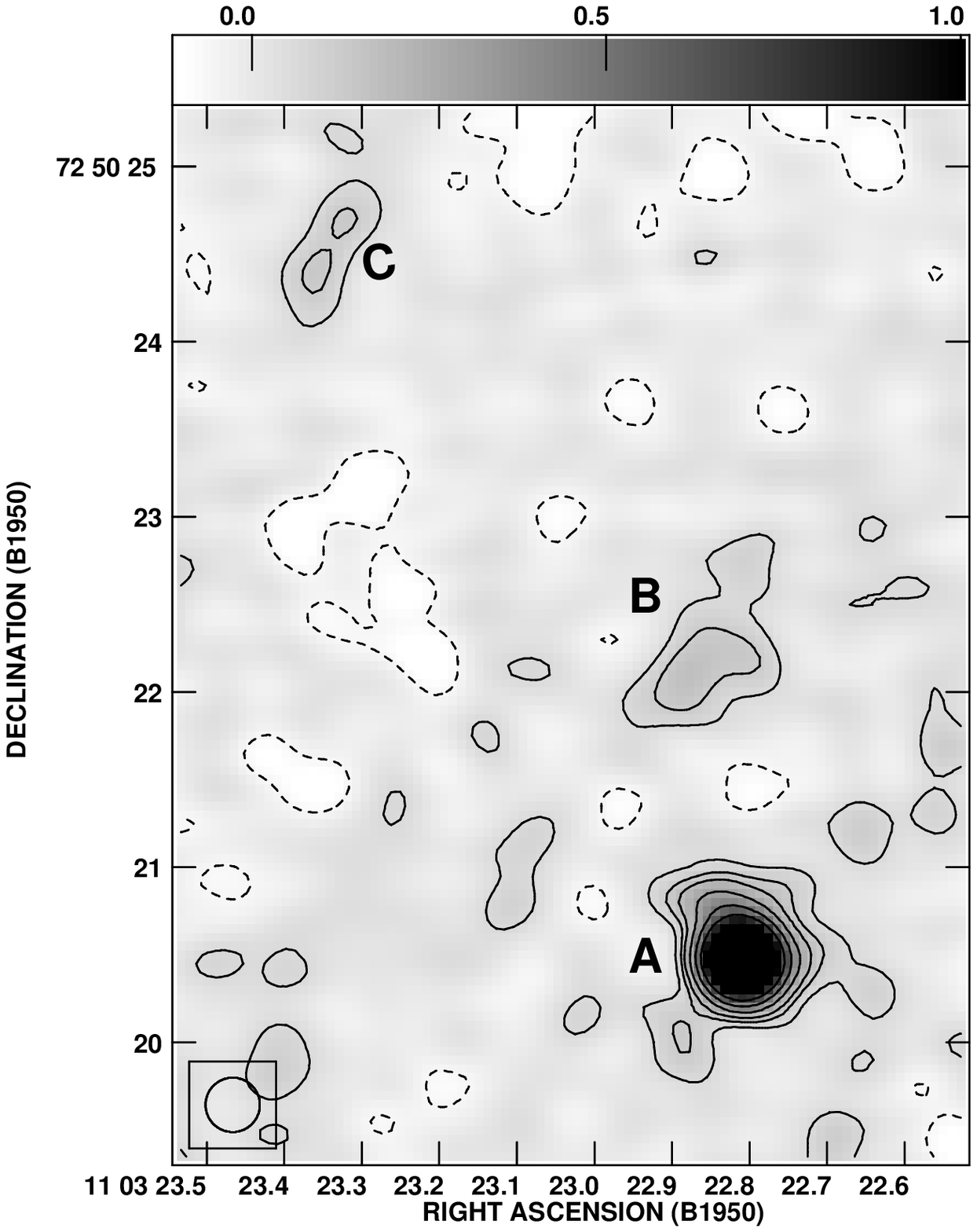} 
\caption{Extended
   emission in NGC~3516; components A, B, C, as identified by Miyaji
   et al (1992) at $\lambda$6 cm and $\lambda$20 cm, are
   labelled. Data were weighted with a 600k$\lambda$ taper in order to
   emphasise these weak components; contour levels are plotted as
   multiples of 1.5-$\sigma$ at (-1, 1, 2, 3, 5, 8, 12, 20, 28,  36)*60~$\mu$Jy beam$^{-1}$.}
\label{fig:tap3516} 
\end{figure}

{\em {\object Mrk~620} (NGC~2273) - Seyfert type 2} - The radio
emission from Mrk~620 (Figure \ref{fig:Seyferts2}) consists of an
east-west double source with weaker, extended emission to the
north-east and south-west (see also Figure \ref{fig:m620sub} :left). The
position of the optical nucleus is close to the eastern component,
suggesting it is the nuclear radio component (N99). However, it is
also possible that the nucleus lies between the two radio sources. The
east component has a slightly flatter spectrum, derived from
matched-resolution MERLIN (1.4 GHz) and VLA (8.4 GHz) images
($\alpha_{1.4}^{8.4}$(E)~$\sim$0.6, where S$_\nu \propto
\nu^{-\alpha}$) than the western component
($\alpha_{1.4}^{8.4}$(W)~$\sim$0.7). However, significant extended
flux is also present, making a reliable estimate of spectral index
difficult. The eastern component shows a weak extension 0\farcs4 to
the east and a stronger western extension reaching $\sim$0\farcs8 and
connecting to the western component of the double (Figure
\ref{fig:Seyferts2}). Gaussian fitting to the eastern component shows that
it is unresolved perpendicular to the source axis, but is extended
along P.A.~81$^{\circ}$; this extended emission is clearly visible in
Figure \ref{fig:m620sub}(right), in which emission from the unresolved point
source has been removed. Table \ref{tab:components} lists measured
positions and flux densities for the eastern (probably nuclear) and
western components; the nuclear flux density shows no variation
between 1992 and 1999 (Table \ref{tab:components}).

The {\em HST} [O{\sc iii}] image \citep{fe00} shows high-excitation
gas with a one-sided jet-like structure extending along the same
east-west direction as the radio emission, along with fainter [O{\sc
iii}] emission extending northeast and southwest of the nucleus. Poor relative
astrometry between {\em HST} optical and radio reference frames makes exact
registration of radio and optical structures uncertain; if the eastern
radio peak is assumed to be the nucleus and aligned with the [O{\sc
iii}] peak, as was done in Figure \ref{fig:m620sub}(left), the [O{\sc iii}]
jet-like feature extends east of the nucleus, where little extended
radio emission is detected, and the radio emission extends primarily west
of the nucleus, where little high-excitation optical emission is
detected. Alternatively, the western radio component could coincide
with the peak in the [O{\sc iii}] emission (resulting in the eastern
radio component being $\sim$0\farcs6 east of the [O{\sc iii}] peak),
suggesting that the [O{\sc iii}] peak could be excited by the impact of the
radio jet on NLR gas.
 
In addition to the east-west double radio emission, we estimate a
further $\sim$3.9 mJy of flux to be present in the diffuse extended
radio emission seen to the northeast and southwest of the double source (Figure
\ref{fig:Seyferts2}).  This diffuse gas might represent back flow from
 radio jets inferred to power the east-west double, or a less
 energetic, poorly collimated outflow wind from the nucleus, similar
 to structures seen in NGC~4051 \citep{ku95,ch97}.

{\em {\object MCG~--5-23-16} (ESO~434-G40) - Seyfert type 2} - This source, which
was not observed by N99, is well detected in our data at a level of
70-$\sigma$ (peak flux).  Weak diffuse emission is detected in the
naturally weighted image (Figure \ref{fig:Seyferts2}, left), in which the
integrated flux density is 20\% higher than the peak. A Gaussian fit
to the naturally-weighted image suggests the source, with a size of
320~mas (58 pc), is marginally resolved in PA $\sim$169$^{\circ}$.  [O
{\sc iii}] emission is elongated on either side of the nucleus in
P.A. 40$^{\circ}$ \citep{fe00} and does not appear to be
related to the marginally extended radio emission.

{\em {\object NGC~3081} - Seyfert type 2} - The radio flux density in this
source has declined by 47\% between 1992 and 1999. The
integrated flux density measured in the naturally-weighted image is slightly
higher than the peak, suggesting the presence of extended emission,
and the source appears to be marginally resolved with a deconvolved
size of $\sim$260 mas (44 pc) in P.A. $-$12$^{\circ}$; this P.A. is
similar to the approximate north-south extension of [O{\sc iii}]
emission, but residual imaging artefacts in the radio image make these
structure estimates tentative.

{\em {\object NGC~3516} - Seyfert type 1.2} - The measured total flux
in 1999 compares well with that in 1992 (Table \ref{tab:allgalprops}), but we are able
now able to decompose the emission into two components; 2.5 mJy of
emission is associated with the unresolved nucleus and 1.4 mJy from a
small component $\sim$0\farcs3 north-east of the nucleus. We cannot,
therefore, rule out variability of the nucleus between 1992 and 1999,
but it is more likely that the total flux measured in the 1992 data includes
both components and the nucleus has remained constant over this
period. As described in Section 2, NGC~3516 was observed in spectral
line mode to reduce bandwidth smearing effects and allow the removal
of contaminating emission from a strong confusing source 4\farcm3
away, which reduced the imaging capabilities of N99; this technique
worked well, although low level contamination is still present
resulting in a measured rms in the final images that is slightly
higher than the theoretical expectation. To test the robustness of our
imaging, a heavily tapered image was created; as can be seen in Figure
\ref{fig:tap3516}, our 8.4 GHz image compares well with previous
images made at $\lambda$20 cm and $\lambda$6 cm by Miyaji et
al. (1992) - components A, B and C, following the nomenclature of
Miyaji et al. (1992) are clearly visible.  The close correspondence
between optical and radio structures in the inner $\sim$4\arcsec\ in
NGC~3516 is discussed in detail by Ferruit et al. (1998).

{\em {\object NGC~4117} - Seyfert type 2} - NGC~4117 was not detected at 8.4 GHz
in 1992 by N99, but is clearly detected in our 1999 data (Figure
\ref{fig:Seyferts3}). Following careful reprocessing of the 1992 data, we confirm that this source is undetected and that atmospheric decorrelation is not responsible for its non-detection. We detect no emission above a 3-$\sigma$ level of 90$\mu$Jy in the 1992 data (providing a very conservative upper limit $<0.1$~mJy); in contrast, we detect a peak flux density at a $\sim$15-$\sigma$ level of
$\sim$0.25 mJy (Figure \ref{fig:Seyferts3}) in the 1999 data. The source is unresolved and still easily detected in the noisier uniform image (Figure \ref{fig:Seyferts3}).

The unresolved radio emission is consistent with the compact [O {\sc
iii}] emission. The more extended, lower excitation [N {\sc
ii}+H$\alpha$] emission \cite{fe00} seems to correspond to
patches of non-nuclear radio emission seen in the 1.4-GHz image of N99
and is likely to be related to star-formation rather than AGN processes.

{\em {\object NGC~4968} - Seyfert type 2} - No variability in the nuclear flux density
of NGC~4968 was detected between 1992 and 1999; the slightly higher total flux
density (7.8~$\pm$~0.7 mJy) listed for the source in 1999 in Table
\ref{tab:allgalprops} includes the nuclear flux density of 6.1
mJy (consistent with the peak nuclear flux in the 1992 data of 6.2$\pm$0.3 mJy) and an additional 1.7 mJy of emission from a component extending
up to $\sim$1$''$ west of the nucleus (Figure \ref{fig:Seyferts3}).  This
extended component was tentatively identifed by N99 and is confirmed
here; it does not appear to correspond to any ionised gas feature,
although dust obscuration west of the nucleus might explain the lack
of detectable optical emission in this region \citep[see Figure 14 in][]{fe00}.

{\em {\object MCG~--6-30-15} (ESO~383-G35) - Seyfert type 1.2} - The
flux density of this source, measured to be 0.64 mJy in our naturally-weighted
image (Figure \ref{fig:Seyferts4}), appears to have declined by $\sim$36\%
since 1992.  However, the integrated flux density measured in the
naturally-weighted image is 1.8 times greater than the peak,
suggesting the presence of extended emission; the source appears to be
marginally resolved with a deconvolved size of $\sim$610~mas in
P.A. 0$^{\circ}$. The source is also clearly detected in the
uniformly-weighted image (Figure \ref{fig:Seyferts4}) with a peak flux
density of 0.33 mJy (i.e. 14 $\sigma$).  A low resolution tapered image
was made, using a 400-k$\lambda$ taper, to search for additional
extended emission; the integrated flux density in the tapered image
(not shown) is 0.81$\pm$0.07~mJy, close to but still lower than the value measured in 1992. The low declination of this source results in a highly elongated beam, particularly for snap-shot observations, and may
account for N99's conclusion that this source is unresolved.

Extended nuclear [O {\sc iii}] emission in MCG~--6-30-15 has been
attributed to emission from an inclined nuclear disk \citep[their
Figure 17]{fe00} and therefore is not expected to have any
associated non-thermal radio emission. Although the structure of the 
extended radio emission is not well determined, our data suggest a
possible elongation approximately perpendicular to the [O {\sc iii}]
disk, as would be expected from a jet-like or disk-driven wind
component. Future sensitive, high-angular resolution observations of this source are required to separate extended and compact radio components and to determine the nature of the nuclear flux density variability.
 
{\em {\object NGC~5273} - Seyfert type 1.5} - The nuclear flux density of this
source has declined by $\sim$40\% between 1992 and 1999 (Table
\ref{tab:allgalprops}).  Gaussian fits by N99 suggested the presence of
some marginally extended emission in P.A. $\sim$170$^{\circ}$ and
Ulvestad \& Wilson suggested P.A. $\sim$5$^{\circ}$. We confirm the
presence of extended emission. A Gaussian fit to our
naturally-weighted image (Figure \ref{fig:Seyferts4}) yields an integrated
flux density that is 1.6 times greater than the peak, with a
deconvolved size of 240~$\times$~133~mas extended in
P.A.~$\sim$177$^{\circ}$, consistent with the north-south extension
seen in ionised gas. Emission in the uniform image is marginally
resolved in P.A. $\sim$167$^{\circ}$ (Figure \ref{fig:Seyferts4}).

{\em {\object NGC~7465} (Mrk~313) - Seyfert type 2} - Our measured total flux of
1.2 mJy is consistent with that measured in 1992 (Table \ref{tab:allgalprops}),
indicating no variability between 1992 and 1999. As can be seen in the
naturally-weighted image (Figure \ref{fig:Seyferts4}), and also tapered
images (not shown), this weak source consists of an unresolved nucleus
and extended emission to the north-east and possibly
south-west. Gaussian fitting to the naturally weighted image suggests
a nuclear source size of 300$\times$200 mas (42$\times$28~pc) in
P.A. 33$^{\circ}$ with a flux density of $\sim$0.8 mJy, with an
additional 0.4~mJy in extended emission.  The higher resolution
uniform image shows a curved structure which suggests extension along
this same P.A., although Gaussian fitting suggests a P.A. closer to
20$^{\circ}$.  The exact decomposition of extended and compact
emission fluxes is uncertain (Table \ref{tab:components}) making it difficult to completely rule out nuclear variability; future observations of this source may provide confirmation of its nuclear behaviour.

Regions of high excitation optical line emission in the inner 2\arcsec\
\citep[Figure 25]{fe00} are elongated in the southeast-northwest
directions, roughly perpendicular to the extended radio emission.

\section{DISCUSSION}

The primary power source for nuclear emission from radio-quiet AGN has
long been a matter of debate, but evidence is mounting in favor of
accretion by a central supermassive black hole over compact nuclear
starbursts.  In particular, the presence of well-collimated radio jets
in some sources provides strong support for the central black hole
plus accretion disk model \citep{ul99,ku99,mn03}, with the accretion
flow and jet forming a natural symbiosis \citep{fb95,fb99}; detection
of radio nuclei with flat radio spectra and high brightness
temperatures (T$_{\rm B}$~$>$~10$^8$~K) in Seyferts \citep{mn00},
LLAGN \citep{N02} and radio-quiet quasars \citep[RQQs;][]{bb98} further
supports this paradigm.

Nuclear flux variability can provide important additional constraints
on central engine physics, but few surveys of radio variability in
radio quiet AGN, particularly Seyferts, exist.  Falcke et al. (2001)
conducted a study of 8.4-GHz radio emission from 30 radio-quiet and
radio-intermediate quasars selected from the PG quasar sample and
found that $\sim$80\% of the sources show at least marginal evidence
for variability over a two year period, with some sources showing
significant month-to-month variability and a possible trend for greater
variability to be seen in nuclei with more inverted spectra. These
results support an AGN origin for the radio emission and suggest the
presence of relativistic parsec-scale radio jets in the RQQs with the
strongest intra-year variability.

\subsection{Radio Variability, Seyfert Type and Nuclear Structure}

As described in Section 3, five out of eleven Seyferts in our sample
show nuclear flux variation over a seven year period. We find no
correlation between the detection or degree of variability and Seyfert
type: although the sample is small and dominated by Seyfert type 2s,
variability is detected in three out of eight type 2 Seyferts and two
out of three Seyfert 1.2/1.5s.   In radio-loud
objects, radio variability is strongest in objects viewed along the
line of sight close to the direction of a relativistic jet (blazars,
OVV quasars). In the context of the unified scheme, such objects would
be of type 1. However, proper motion measurements of Seyfert radio
components on the parsec and sub-parsec scale reveal non-relativistic motions
\citep[e.g.][]{ul99,ro00,mi04,ul05}, so amplification by bulk relativistic
motion may not occur. Thus the physical nature of nuclear radio
variability may be different in radio-loud and radio-quiet AGN.  A
study of a larger sample of Seyferts, less dominated by type 2s, is
required to investigate further whether there is any difference
between the radio variability properties of type 1 and type 2 Seyferts.

There is also no correlation between the detection of variability and
core luminosity, which lies in the range
8.2~$\times$~10$^{18}$~$<$~L(core)$_{\rm 8.4~GHz}$~$<$~
5.9~$\times$~10$^{21}$~W~Hz$^{-1}$ for this sample, but we do find a
correlation between variability and radio compactness. All variable
sources, except NGC~2110, are either unresolved or only tentatively
marginally resolved, while flux densities of core components in
resolved or linear sources are constant at the two epochs.  This
suggests that all Seyferts might exhibit variation in their nuclear
radio flux density at 8.4 GHz, but that variability is more easily
recognized in compact sources in which emission from the variable
nucleus is not mixed with unresolved, constant flux density radio-jet
emission within the central $\lesssim$~50 pc (i.e. the $\sim$0\farcs2
typical VLA beam size for V$_{sys}$$<$4000~km~s$^{-1}$). The increased
detection rate of flat spectrum nuclear radio emission in Seyferts
imaged with VLBI resolution \citep{mi04} further
emphasises the importance of isolating nuclear emission from jet
emission. Linear resolution limitations for studies of more distant
AGN such as RQQs are more accute. Barvainis et
al. (2005) find, on average, few differences between radio-loud quasars (RLQs) and
RQQs other than an unexpectedly-weak dependence of
radio variability on nuclear spectral index, with RQQs showing on
average steeper spectral indices than RLQ. If RQQs have
extended, but unresolved steep spectrum emission similar to that seen
in Seyferts, their nuclear variability and nuclear spectral index
might also have been underestimated, thus further increasing their
similarity to radio-loud AGN.

\subsubsection{Nuclear Flare in NGC~2110?}

NGC~2110 is unusual as the only Seyfert in this sample with
significant extended radio emission {\em and} a variable nucleus,
suggesting that the intrinsic nuclear variability may be higher than
the observed $\sim$38\%.  As can be seen in Figure \ref{fig:Seyferts1},
NGC~2110 has striking extended radio structure in the form of curved
symmetrical radio jets extending $\sim$4\arcsec\ north and south of a
compact central flat-spectrum nucleus \citep[see
also][]{uw83,uw84,N99}. On the smallest scales, Mundell et al. (2000)
identified the nucleus at 8.4~GHz as an 8-mJy point source with a
maximum size of 0.14~$\times$~0.07 pc and brightness temperature in
excess of 6.0~$\times$~10$^8$~K, consistent with sychrotron self
absorption and a black-hole driven central engine. They also resolved
extended emission $\sim$0.7~mas (0.1 pc) north and south of the
nucleus consistent with emission from the inner regions of the
northern and southern jets, and a discrete component $\sim$1.95 mas
($\sim$0.3 pc) north of the nucleus, possibly the first knot in the
northern jet.  The nuclear flux variability measured with the VLA
clearly indicates a significant change in the structure of the nucleus
on scales smaller than the VLA beam size, within the central
$\sim$0\farcs1 (15 pc), between the two epochs, possibly due to fading
of an earlier flare in which new components or shocks in the jet may
have appeared.  Multi-epoch, multi-frequency VLBA imaging of the
nuclear regions would establish the cause of the variability and
determine whether new components are emerging in the jet flow during
periods of increased nuclear flux density.

\subsection{Nuclear Flares and Radio Jets}

We are unable to quantify the minumum timescale of variation due to
the relatively long time between our two epochs of 8.4-GHz
observations. However, the serendipitous discovery of variation in
$\sim$50\% of our sample, which was not selected on any given radio
property or known variability, suggests that radio variability in
Seyfert nuclei may be a common phenomenon and that the measured amplitudes
of variation are likely to be lower limits to peak-to-peak variations
within this period.  

In radio-loud objects, the bulk relativistic flow in powerful radio
jets is thought to play a key role in Doppler boosting the amplitude
of nuclear flux variation and compression of variation timescales,
thus enhancing the observed flux variability in blazars whose jets are
viewed close to the ejection axis.  Long-term radio monitoring of
radio-loud AGN \citep[e.g.][]{hu92,va92} has demonstrated the
existence of stochastic, non-random flare events at centimeter and
millimeter wavelengths, with mean timescales of 1.95 and 2.35 years
for BL Lac objects and QSOs respectively \citep[e.g.][]{hu92}.  In
standard jet-shock models, the increases in total flux density are
attributed to the passage of shocks on parsec-scales in collimated,
relativistic flows and, in some cases, outbursts in monitoring data
are seen to correspond to propagation of individual components in VLBI
maps.

In contrast, the debate continues over the nature of the ejected radio
plasma in Seyferts; the presence of bulk relativistic motion has not
yet been demonstrated in Seyfert jets, with existing proper motion
studies measuring apparent radio component speeds
$V_{app}$~$\lesssim$~0.25~c \citep{ul99,ro00,ro01}. The observations
of one-sided parsec-scale jets can be explained by free-free
absorption by ionized gas in a disk or torus along the line of sight
to the counterjet, rather than Doppler de-boosting/dimming
\citep{ul99}. Nevertheless, despite the lack of systematic long-term
monitoring of Seyferts, radio flares have been detected in a small
number of Seyfert nuclei.

A 10-year monitoring study of the 8.4-GHz nuclear radio emission in
the double radio-lobed Seyfert 1 galaxy NGC~5548 \citep{wr00}
revealed photometric variations of 33\% and 52\% between VLA
observations separated by 41 days and 4.1 years respectively. In
addition, during the 41-day flare the nucleus exhibited an inverted
spectrum with spectral index $\alpha$~$\sim$~-0.3 (S$_\nu \propto
\nu^{-\alpha}$) compared with the steeper value $\alpha$~$\sim$~0.2
prior to the flare. The flare characteristics support a black-hole
driven central engine for the origin of the radio emission with the
spectral inversion possibly arising from brightening in the
optically-thick base of the parsec-scale radio jet; efforts continue
to detect a secondary ejected component.

More dramatic radio flares and inverted radio spectra have been
detected in Mrk~348 and III~Zw~2 \citep[e.g.][]{bh05}.  The variable
compact radio nucleus of the Seyfert 2 galaxy Mrk~348
\citep[e.g.][]{sr74,nb83} showed increases in continuum flux density
by factors of 1.7 at 5 GHz and 5.5 at 15 GHz during the
1982.3$-$1983.3 and 1997.10$-$1998.75 periods respectively
\citep{nb83,ul99}; bright H$_2$O megamaser flares were also detected
in 2000 and 2001 \citep{fa00,pe01}.  In III~Zw~2, the high radio
frequency light curves (22 GHz and 43 GHz) measured over a $\sim$20
year period show a number of radio flares; of particular note was the
flare in 1998 that showed a $\sim$30-fold flux increase within two years
and a highly-inverted radio spectrum peaking at millimeter wavelengths
\citep{al85,fa98}; similar flares are evident at earlier epochs over
the last $\sim$25 years at lower frequencies \citep{bh05}. A sudden
change in spectral shape in 1999, with the frequency of the flux peak
dropping quickly from 43 GHz to 15 GHz within a few months, indicated
a structural change in the nucleus; this was confirmed by high
resolution imaging of the nucleus that revealed a newly ejected radio
component expanding at an apparent superluminal speed
$V_{app}$~$\gtrsim$~1.25~c \citep{bh00}, in stark contrast to the
upper limit of 0.04~c measured during quiescence. Brunthaler et
al. (2003) suggest that III~Zw~2 flares approximately every five
years; this timescale is consistent with the variability detected for
Seyferts in our sample, given that two-epoch observations are more
likely to detect fading sources if the decay timescale is longer than the brightening timescale, but shorter timescale variations,
such as those seen in quasars \citep{bv05} cannot be ruled out.

\subsubsection{Origin of Radio Flares: Jet-Cloud Interactions?}

Due to the small number of detailed studies of Seyfert nuclei in radio
outburst, particularly the lack of parsec to sub-parsec scale imaging,
the origin of such flares is not yet known. It has been suggested that
shocks produced by interaction of a relativistic jet with dense
circumnuclear gas clouds within the central parsec might be
responsible for the flares, with in situ particle re-acceleration in
the shock producing brightening radio emission and inverted spectra
\citep{fa99}. For III~Zw~2, the initial flux density increase
and inverted spectrum with a low jet speed have been likened to
similar properties of Gigahertz-Peaked-Spectrum (GPS) sources, in
which ultra-compact hot spots are powered by a relativistic jet
interacting with circumnuclear clouds \citep{bh03}. The
subsequent observed phase of superluminal expansion might then
correspond to the jet breaking free and propagating relativistically
into a lower-density medium \citep{bh00}.

Water vapour megamaser flares may also be consistent with a jet-cloud
interaction scenario in Seyfert nuclei. Falcke et al. (2000) argued
that the megamaser flare in Mrk~348 was driven by a corresponding
increase in the continuum emission and suggested the line responded to
the continuum flare within $\sim$2-years. The upper limit to the line
response time implies the masers lie within $\lesssim$~0.6~pc of the
nucleus, a size confirmed by direct, high-resolution imaging (Peck et
al. 2001). This imaging showed the maser emission lies $\sim$0.4~pc
north of the nucleus towards the northern jet, rather than in a
nuclear disk perpendicular to the jet.  Large H$_2$O linewidths on
these small spatial scales are consistent with the H$_2$O emission
arising from a shocked interface between energetic jet material and
the intervening molecular gas cloud, a scenario further supported by
spectral evolution mimicking that seen in III~Zw~2, with a shift in
spectral peak to lower energies at later times \citep{pe01}.  NGC~1052
shows similar jet-aligned masers \citep{cl98} consistent with an
impact between the jet and a molecular cloud.

More circumstantially, the bright shock-like features in the radio
continuum structure of the jet in NGC~4151 have been attributed to the
interaction of a well-collimated jet with circumnuclear clouds of
diameter $<$1.4~pc \citep{mn03}; stringent upper limits on
radio component speeds are consistent with the jet being slowed to
sub-relativistic speed on scales significantly smaller than $\sim$3~pc
\citep{ul05}.

\subsection{Intermittent Nuclear Quiescence and Outburst}

Key outstanding questions remain on the speeds of Seyfert radio jets,
the nature of radio variability in Seyfert nuclei and the relationship
between the two.  Radio variability and spectral evolution studies
offer valuable tools to probe the nuclear radio emission in large
numbers of Seyfert nuclei, before following up objects in outburst
with high resolution, VLBI imaging. The ejection speeds of Seyfert
jets may turn out be intrinsically non-relativistic. However, the low
radio component speeds measured in the early phase of the flare in
III~Zw~2 might have important implications for interpretation of
sub-relativistic speeds measured in other Seyferts, such as Mrk~348
and Mrk~231 \citep{ul99}, in which the low speeds may
indicate observations taken during nuclear quiescence or early
outburst stages when the jet is impacting a cloud but has not yet
drilled through.  The small detection rate ($\sim$7\%) of H$_2$O megamaser
emission in nearby AGN \citep[e.g.][]{br96} and
 maser variability \citep{br04} may, in
part, result from variability in the 22-GHz continuum, which is
presumed to provide the seed photons for maser amplification, and thus
further supporting the paradigm of intermittent periods of quiescence
and nuclear outburst across the Seyfert population.

Systematic, multifrequency radio monitoring of a larger sample of
Seyfert nuclei over timescales of months to years is required to
determine the characteristic timescales and amplitudes of variation
and changes in spectral index, as well as any correlations with
Seyfert type and radio structure. Rapid follow up of flaring nuclei at
other wavelengths, in combination with radio imaging on parsec and
sub-parsec scales, would provide valuable insight into structural
changes, the possible presence of relativistic flows and their relationship
with accretion disk changes during the outburst. Such observations
will become increasingly routine with new facilities such as the EVLA,
eMERLIN and LOFAR.

\section{Conclusions}

We have used the VLA at 8.4 GHz to study the radio continuum emission
from a sample of 12 optically-selected, early-type Seyfert
galaxies. We find the following:

\begin{itemize}
\item All 12 Seyferts are detected with the VLA; five nuclei show
 a variation in their nuclear flux density since 1992, while six have
 remained constant (one was not observed in 1992).  No correlation is
 found between Seyfert type and the detection of radio variability.
 Instead, the compactness of the radio emission seems to be a good
 indicator of variability. All variable nuclei, apart from NGC~2110,
 are compact and at best marginally resolved, while nuclei with
 associated extended radio emission have remained constant.

\item NGC~2110 is the only Seyfert in this sample with significant
extended radio emission from well-defined radio jets {\em and} a
variable nucleus, suggesting that the intrinsic nuclear variability is
higher than the observed $\sim$38\%.  The observed nuclear flux
variability indicates significant changes are likely to have occurred
in the structure of the nucleus on scales smaller than the VLA
beam size (i.e. within the central $\sim$0\farcs1 [15 pc]), between the
two epochs, possibly due to the appearance and fading of new
components or shocks in the jet. This interpretation is consistent
with sub-parsec scale structures identified previously by Mundell et
al. (2000) that consist of a high brightness temperature
($>$~6.0~$\times$~10$^8$~K), synchrotron self-absorbed point source,
extended jet emission 0.1 pc north and south of the nucleus, and a
discrete component $\sim$0.3 pc north of the nucleus, possibly the
first knot in the northern jet.

\item The serendipitous discovery of variability in $\sim$50\%
of our sample, which was not selected on any given radio property or
known variability, suggests that radio variability in Seyfert nuclei
is a common phenomenon and that the measured amplitudes of variation
are likely to be lower limits to peak-to-peak variations within this
period. 

\item We conclude that all Seyferts might exhibit variations in their
nuclear radio flux density at 8.4 GHz, but that variability is more
easily recognized in compact sources in which emission from the
variable nucleus is not mixed with unresolved, constant flux density
radio-jet emission within the $\sim$0\farcs2 VLA beam size (i.e. the
central $\lesssim$~50 pc for V$_{sys}$$<$4000~km~s$^{-1}$).

\item We speculate that if flares in radio light curves correspond to
ejection of new relativistic components or emergence of shocks in the
underlying flow, sensitive systematic monitoring of a larger sample,
combined with subarcsecond resolution imaging, may confirm that
Seyferts - as black-hole driven AGN - have the capacity to accelerate
relativistic jets during radio flares despite having radio jets that
are intrinsically non-relativistic during quiescence.

\end{itemize}

\acknowledgements
This paper is dedicated to A.S. Wilson, a valued colleague, mentor and friend. 
CGM acknowledges financial support from the Royal Society and Research
Councils U.K; NN acknowledges  support from BASAL PFB-06/2007, ALMA 31060013
 and Fondecyt 1080324. We thank the anonymous referee for constructive comments
that improved the paper. This
research was partially supported by NSF grant AST 9527289 to the
University of Maryland.  The National Radio Astronomy Observatory is a
facility of the National Science Foundation operated under cooperative
agreement by Associated Universities, Inc.  This research has made use
of the NASA Astrophysics Data System Abstract Service (ADS), and the
NASA/IPAC Extragalactic Database (NED), which is operated by the Jet
Propulsion Laboratory, California Institute of Technology, under
contract with the National Aeronautics and Space Administration.

{}
\end{document}